\documentclass[aps,pra,twocolumn,superscriptaddress,showpacs,showkeys,amsmath,amssymb]{revtex4}
\usepackage{amsfonts}
\usepackage{amssymb,amsmath}
\usepackage{mathrsfs}
\usepackage{latexsym}
\usepackage{amsmath}
\usepackage[cp1251]{inputenc}
\usepackage{graphicx}
\usepackage{dcolumn}
\usepackage{bm}
\usepackage{color}

\RequirePackage{ifthen}
\RequirePackage[pdfstartview=FitH]{hyperref}
\begin{document}
	
	\title{Strongly interacting Bose-Fermi mixtures in $4-\epsilon$ dimensions}
   	\author{V.~Pastukhov\footnote{e-mail: volodyapastukhov@gmail.com}}
	\affiliation{Professor Ivan Vakarchuk Department for Theoretical Physics, Ivan Franko National University of Lviv, 12 Drahomanov Street, Lviv, Ukraine}

	\date{\today}

	
	
	\begin{abstract}
	Thermodynamically stable low-temperature phases of the Bose-Fermi mixtures composed of bosons and spinless fermions close to four dimensions are considered. In the regime, where the only boson-fermion two-body interaction is present and tuned to unitary limit, the properties of a system solely depend on the mass and number ratios of constituent atoms. In addition to the phase with the dimers (boson-fermion shallow bound states), we identified one more state of the mixture with the coexistence of fermionic dimers and trimers. The universal physics of these phases, the characteristic feature of is absence of the Bose-Einstein condensate, is discussed.
	\end{abstract}
	
	\maketitle
\section{Introduction}
In contrast to the condensed matter systems, where the collective effects are dominative in forming of their behaviors, dilute gases (both classical and quantum) with the short-ranged interaction between particles can be described by the sequential inclusion of the few-body effects. In this context, the realization of ultracold Bose and Fermi gases in weakly \cite{Anderson_1995,Davis_1995,DeMarco_1999} and strongly interacting \cite{OHara_2002,Navon_2011} regimes opened new possibilities for the experimental detection of the few-body physics and elucidation of its impact on the macroscopic properties. An extreme diluteness of quantum gases together with the short-range character of the two-body potential among particles make these systems a very convenient playground for testing effective field theories \cite{Bedaque_2002,Braaten_2006}. In such conditions, any specific details of the microscopic potential are irrelevant allowing one to consider theories with contact interactions. The Galilean-invariant Lagrangians with a local two-body coupling are non-renormalizable above two spatial dimensions. However, the intrinsic $U(1)$ symmetry, which guarantees the particle number conservation, provides a possibility to treat two-, three- and generally few-body sectors separately. The ultraviolet (UV) divergences appearing in every next sector can be cured by introducing new higher-order local interactions with the cutoff-dependent coupling constant. Then, the UV dependence of couplings is fixed by requiring the observables to be cutoff-independent. This renormalization procedure which is reminiscent of the Wilson-Polchinski \cite{Wilson_1971,Polchinski_1984} renormalization group (RG) scheme, allows to introduce a set of finite `observable' few-body couplings controlling the universal behavior of a system in the many-body limit.

The Bose-Fermi mixtures of bosons and spinless (spin-polarized) fermions are excellent candidates for observing \cite{Stan_2004,Inouye_2004} composite particles in macroscopic systems. The Pauli exclusion principle stabilizes the fermionic mixture against collapse while numerous Fermi surfaces minimize its total energy. Even at weak contact boson-fermion interaction, except a metastable phase \cite{Saam_1969,Viverit_2000} with almost degenerated Fermi gas and almost undepleted Bose condensate, there is a lower energy state \cite{Powell_2005,Watanabe_2008,Marchetti_2008} with coexistence of bosons, atomic and molecular fermions. If the densities of bosonic atoms are lower than those of fermionic ones, the Fermi-Fermi (fermion-molecular) mixture remains stable without boson-boson repulsion. The Bose-Einstein condensation (BEC) transition in this system at weak boson-fermion coupling both at zero \cite{Fratini_2012,Hryhorchak_2023} and finite \cite{Fratini_2010,Manabe_2019} temperatures realizes simultaneously with the collapse instability. Inclusion of the repulsive boson-boson interaction provides \cite{Guidini_2015} stabilization of the Bose condensate. Recently, the quantum phase transition \cite{Ludwig_2011,Bertaina_2013} from the Fermi-polaron state to the molecular phase was observed \cite{Duda_2023} experimentally. Taking into account a weak induced interaction between fermions and molecules necessarily leads to superfluidity characterized \cite{Guo_2023} by crossover from atomic to molecular pairing mechanisms.

So far, only degenerated Fermi gases of fermions and diatomic molecules were mostly prepared \cite{Park_2012,Heo_2012,Cumby_2013,Cao_2023} in the Bose-Fermi mixtures. In Ref.~\cite{Yang_2022} the formation of triatomic molecules in these systems was observed. However, an important feature of the Bose-Fermi mixtures less discussed in the literature is the formation of fermionic trimers at macroscopic atomic populations. In three dimensions, this problem is complicated by the emergence of the Efimov physics \cite{Efimov,Naidon}. Depending on the set of parameters: densities of constituents, the boson-fermion $s$-wave scattering length, and intrinsic length-scale determining an infinite tower of the three-body bound states, the system may include an arbitrary number of coexisting macroscopically-populated universal trimer states. The richest composition is reached at high densities. Since all trimer levels are `distinguishable' fermions, it is energetically preferable to have as many as possible Fermi surfaces in this limit. In the present paper, we focus on a different and somewhat simpler situation in the non-physical fractional dimensions and for mass ratios of the resonantly-coupled bosons and fermions outside the Efimov window \cite{Rosa_2018,Hryhorchak_2023}. When the only parameter characterizing interaction --the boson-fermion $s$-wave scattering length-- diverges, the thermodynamic properties and stability condition for the mixtured state are universal functions of densities and masses of bosonic and fermionic atoms. Generally, there are no simple ways to calculate these universal functions, but close to four dimensions, the $\epsilon$-expansion can be utilized.

\section{Model and review of $\epsilon$-expansion}\label{}
\begin{figure}[h!]
	\centerline{\includegraphics
		[width=0.20
		\textwidth,clip,angle=-0]{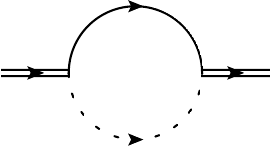}}
	\caption{Self-energy insertion in the dimer propagator (double solid line). Solid and dotted lines denote fermionic and bosonic propagators, respectively.}\label{Self_energy_fig}
\end{figure}

The discussed model describes spinless bosons and spin-polarized fermions in the $d$-dimensional volume $L^d$ with periodic boundary conditions. The interaction between bosonic atoms is neglected, while we assume presence of a contact boson-fermion (pseudo)potential characterized by a bare coupling constant $g_{\Lambda}$ that depends on the ultraviolet cutoff $\Lambda$ 
\begin{eqnarray}\label{g_def}
g^{-1}_{\Lambda}=g^{-1}-\frac{1}{L^d}\sum_{|{\bf p}|<\Lambda}\frac{2m_{bf}}{p^2}.
\end{eqnarray}
For a positive renormalized (`observable') couplings
\begin{eqnarray}\label{g}
g^{-1}=-\frac{\Gamma(1-d/2)}{(2\pi)^{d/2}}m_{bf}^{d/2}|\varepsilon_{c}|^{d/2-1}.
\end{eqnarray}
this model supports a single two-body bound state in a vacuum with the energy $\varepsilon_{c}=-\frac{1}{2m_{bf}a^2}$ (we adopt notation for the boson-fermion reduced mass $m^{-1}_{bf}=m^{-1}_{b}+m^{-1}_{f}$, with $m_b$ and $m_f$ being masses of bosonic and fermionic atoms, respectively; in the following, all $m$s with a single index stand for mass of atoms or composite particles, while the ones with double indices for the reduced masses). The Euclidean action that explicitly incorporates the composite fermions can be obtained by the Hubbard-Stratonovich transformation of the original boson-fermion interaction introducing auxiliary Grassmann fields
\begin{eqnarray}\label{S}
S=\int_xb^{\dagger}\left\{\partial_{\tau}-\xi_b\right\}b+\int_xf^{\dagger}\left\{\partial_{\tau}-\xi_f\right\}f\nonumber\\
+g^{-1}_{\Lambda}\int_xc^{\dagger}c-\int_x\left\{c^{\dagger}fb+b^{\dagger}f^{\dagger}c\right\},
\end{eqnarray}
(with $\int_x$ denoting integration over the $(d+1)$-dimensional Euclidean space-time) where $\xi_{b(f)}=-\frac{\nabla^2}{2m_{b(f)}}-\mu_{b(f)}$ are the vacuum dispersions shifted on the chemical potentials $\mu_{b(f)}$ referring to bosons (fermions). The complex-valued fields $b(x)$, and the Grassmannian fields $f(x)$ and $c(x)$ describe bosons, and two types of fermions: atoms and composite fermions (dimers). As it was first mentioned by Nussinov and Nussinov \cite{Nussinov_2006}, exactly in $d=4$ dimers do not interact mutually and with other particles. This is most easily demonstrated by calculating the vacuum $c$-particle propagator in $(d+1)$-momentum space \cite{Nishida_2012}. The only self-energy contribution is given by a single diagram depicted in Fig.~\ref{Self_energy_fig}
\begin{eqnarray}\label{c_P}
\langle c^{\dagger}_Pc_P\rangle=\frac{g|\varepsilon_{c}|^{d/2-1}}{|\varepsilon_{c}|^{d/2-1}-\left[\frac{p^2}{2m_c}-i\nu_p
		\right]^{d/2-1}},
\end{eqnarray}
with $m_c=m_b+m_f$. Exactly in $d=4$, the above denominator reproduces the inverse propagator of a free composite particle with binding energy $\varepsilon_{c}$. More importantly that arbitrarily close to four dimensions, the residue of the dimer propagator disappears linearly in $\epsilon=4-d$ (see \ref{g}). This fact allows to build the perturbation theory even for strongly interacting systems ($g^{-1}=0$) using $\epsilon$ as a small parameter. Indeed, every diagram containing one dimer propagator and free of UV divergences carries a small parameter. Therefore, all diagrams can be classified by the number of $\langle c^{\dagger}c\rangle$-lines, or equivalently, by rescaling the $c$-fields, we can associate the factor $\sqrt{\epsilon}$ with every interaction vertices in action (\ref{S}). All observables then are calculated as a power series in $\sqrt{\epsilon}$. This was shown \cite{Nishida_2006} to be an extremely effective analytical tool for the calculation of thermodynamic and spectral properties of spin-$1/2$ fermions at unitarity. In the rest of the paper, the $\epsilon$-expansion will be used for examining properties of Bose-Fermi mixtures mainly in the limit of infinite boson-fermion coupling. In the two-body sector, this limit of our model is characterized by the scale (and even conformal) invariance. If the three-body sector is also trivial, i.e. there is no Efimov effect in the considered system, all the above symmetries are preserved.

\section{Dimer-fermion mixture}\label{}
\begin{figure}[h!]
	\centerline{\includegraphics
		[width=0.475
		\textwidth,clip,angle=-0]{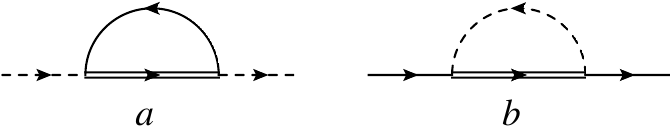}}
	\caption{Bosonic (a) and fermionic (b) self-energies of order $\epsilon$.}\label{bf_self_en_1_fig}
\end{figure}

By neglecting boson-boson interaction we restrict our consideration to a case of the non-Bose-condensed ground states of the system, i.e. no single-particle energy level is macroscopically occupied by the Bose atoms. The presence of BEC necessarily leads to the collapse of the mixture. This means that all bosons should be bound to fermionic atoms forming dimers. It is easy to show that there is always a region in the parameter space, where the mixture state of the system close to $d=4$ at unitarity is stable. Indeed, in this limit, we are dealing with two ideal Fermi gases of fermions and dimers. The number of dimers is equal to the number of bosons, while all other fermions are free
\begin{eqnarray}\label{dimer_Eqs}
n_b=\frac{1}{L^d}\sum_{{\bf p}}\theta(-\xi_c(p)),\\
n_f-n_b=\frac{1}{L^d}\sum_{{\bf p}}\theta(-\xi_f(p)),
\end{eqnarray}
where $\theta(x)$ is the Heaviside step function and $\xi_c(p)=\frac{p^2}{2m_c}-\mu_c$ with $\mu_c=\mu_b+\mu_f$ being the chemical potential of composite particles. By finding $\mu_f$ and $\mu_b$ one can demonstrate the thermodynamic stability of mixture, i.e. $\frac{\partial \mu_f}{\partial n_f}>0$, $\frac{\partial \mu_b}{\partial n_b}>0$ and $\frac{\partial \mu_f}{\partial n_f}\frac{\partial \mu_b}{\partial n_b}-\frac{\partial \mu_f}{\partial n_b}\frac{\partial \mu_b}{\partial n_f}>0$ for all
\begin{eqnarray}\label{dimerBEC_cond}
\frac{n_f}{n_b}\ge 1+\left(\frac{m_f}{m_c}\right)^{d/2}.
\end{eqnarray}
For lower densities of fermions, the chemical potential of bosons crosses zero ($\mu_b<0$ in the thermodynamically stable region) signaling the BEC transition. Recall that the above conclusions are true only for $d$ extremely close to four. At small but non-zero $\epsilon$ we can elucidate properties of the system perturbatively. Since there are no free bosons, the system is a weakly interacting Fermi-Fermi mixture. Because of the fermionic statistics, this interaction is repulsive in the $s$-way channel and one can apply the Fermi liquid paradigm to calculate the particle distribution and thermodynamics of the system. The interaction in such a system only deforms the form of the momentum distribution but does not the position of its discontinuity point. By computing the self-energies of the unbound fermion (Fig.~\ref{bf_self_en_1_fig}, $b$) 
and dimer propagators, therefore, we can identify the chemical potentials $\mu_f$ and $\mu_c$, while correction to the bosonic correlator (Fig.~\ref{bf_self_en_1_fig}, $a$) modifies the BEC transition condition (\ref{dimerBEC_cond}). Lengthy, nonetheless straightforward calculations lead to inequality (up to order $\epsilon$)
\begin{eqnarray}\label{dimerBEC_cond_eps}
&&\frac{n_f}{n_b}\ge 1+\left(\frac{m_f}{m_c}\right)^{d/2}\nonumber\\
&&+\epsilon\frac{m_f}{m_b}\left[\frac{m_c}{m_b}+\frac{m_f}{m_c}-\frac{m_c}{2m_f}-\frac{m^2_f}{2m^2_c}\right],
\end{eqnarray}
for the thermodynamic stability. Note that for a strong imbalance between masses of bosonic and fermionic atoms $m_b\ll m_f$, the correction can be arbitrarily large shifting the stability condition towards larger fermion densities.

\begin{figure}[h!]
	\centerline{\includegraphics
		[width=0.475
		\textwidth,clip,angle=-0]{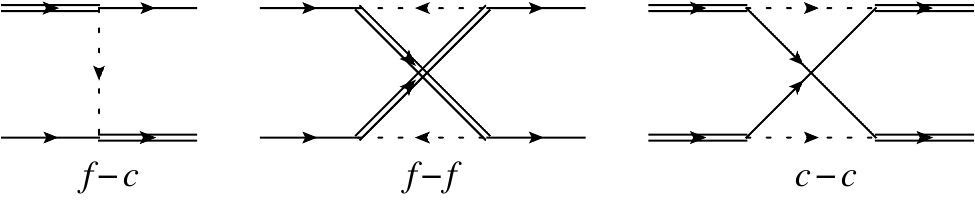}}
	\caption{Leading-order effective fermion-dimer ($f-c$), fermion-fermion ($f-f$) and dimer-dimer ($c-c$) interactions.}\label{Superfluid_vertices_fig}
\end{figure}

\begin{figure}[h!]
	\centerline{\includegraphics
		[width=0.475
		\textwidth,clip,angle=-0]{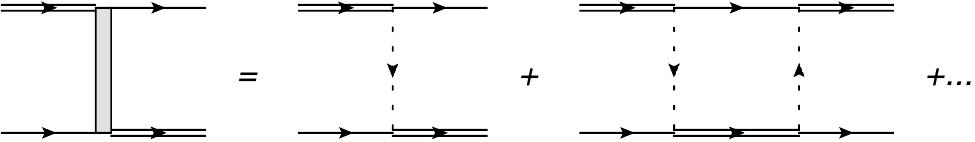}}
	\caption{A sum of the ladder diagrams determining the exact fermion-dimer vertex.}\label{Sum_for_superfluid_fig}
\end{figure}

It is well-known that a macroscopic number of fermions are unstable towards the Cooper pairing at attractive interaction of any magnitude. In our case, when two sorts of Fermi particles (unbound fermions and dimers) are built of the same fermionic atom, the only possible pairing channel is the $p$-wave one. There are, however, a few possible scenarios of the Cooper pairing compositions, namely, atom-dimer $f-c$, atom-atom $f-f$, and dimer-dimer $c-c$ (and all possible phases with the coexistence of different pairings). The latter two are less energetically preferable since the appropriate interactions are of order $\epsilon^2$ (see, Fig.~\ref{Superfluid_vertices_fig}). 
The atom-dimer effective interaction is of order $\epsilon$, and moving from the normal phase towards lower temperatures, the $p$-wave $f-c$ superfluidity should be observed. In order to study the peculiarities of this pairing mechanism to leading order in $\epsilon$, one needs to sum up an infinite series of ladder diagrams presented in Fig.~\ref{Sum_for_superfluid_fig}. 
The series summation is equivalent to solving an integral equation for the two-body fermion-dimer vertex function \cite{Levinsen_2011,Brodsky_2006}. Being interested in the $f-c$ bound states we only need to find a non-trivial solution to the homogeneous part of this vertex
\begin{eqnarray}\label{Z_P}
Z_{fc}(P)=\frac{1}{\beta L^d}\sum_{Q}\langle b^{\dagger}_{P+Q+K}b_{P+Q+K}\rangle\langle f^{\dagger}_{Q}f_{Q}\rangle\nonumber\\
\times\langle c^{\dagger}_{Q+K}c_{Q+K}\rangle Z_{fc}(Q),
\end{eqnarray}
where $i\omega_k$ after analytic continuation should be identified with binding energy and ${\bf k}$ with the $d$-dimensional wave-vector of the fermion-dimer Cooper pair. In general, the solution is complicated but close to 4D, where the dimer propagator (\ref{c_P}) acquires a pole structure with vanishingly small residue, Eq.~(\ref{Z_P}) formalizes in the system of coupled linear integral equations for functions $\psi_{\bf p}=Z_{fc}(P)_{i\nu_p\to -\xi_f(p)}$ and  $\phi_{\bf p}=Z_{fc}(P)_{i\nu_p\to \xi_c(|{\bf p}+{\bf k}|)-i\omega_k}$. The first one is determined for all $|{\bf p}|>p_f$, while the second one is only non-zero for all $|{\bf p}+{\bf k}|<p_c$. Considering for simplicity only case of $p_f=p_c$ (otherwise, the Cooper pairing requires a finite fermion-dimer momentum) and ${\bf k}=0$, one can argue that  $\psi_{\bf p}={\bf n}{\bf p}\psi_p$ and $\phi_{\bf p}={\bf n}{\bf p}\phi_p$. It is possible to find the large-$p$ behavior (scaling limit) of $\psi_p\sim 1/p^{\sigma}$ in arbitrary dimension with exponent $\sigma$ given by equation
\begin{eqnarray}\label{sigma}
&&\frac{m_c}{m_f}=\frac{\sigma-2}{d}\frac{\sin(\pi d/2)}{\sin(\pi\sigma/2)}\nonumber\\
&&\times _2F_1\left(1+\frac{d-\sigma}{2},\frac{d+\sigma}{2}-1; \frac{2+d}{2};\frac{m^2_f}{m^2_c}\right),
\end{eqnarray} 
[here $_2F_1\left(\dots\right)$ is the hypergeometric function \cite{Abramowitz}] which is nothing but the scaling dimension of the appropriate $p$-wave channel ($fc$) composite operator \cite{Nishida_2012}. Note that the solutions to Eq.~\ref{sigma} go in pairs $\sigma$ and $4-\sigma$. In the limit $\epsilon\to 0$ exponent $\sigma$ linearly decreases to zero. For small enough masses of bosons, there is a region with complex conjugated exponents $\sigma=2\pm i\sigma_0$ of the $p$-wave Efimov effect emergence at arbitrary $d$. Extremely close to $d=4$, the effective fermion-dimer interaction almost vanishes, which allows to approximate functions $\psi_p\approx\psi_{p_f}$ and $\phi_p\approx\phi_{p_f}$ by their values at Fermi surface (recall that here we only discuss the limit of equal fermion and dimer populations). This simplification transforms two integral equations (for $\psi_p$ and $\phi_p$) into a single algebraic one with the non-trivial solution determining with the exponential precision, the fermion-dimer binding energy (and consequently gap in the quasi-particle spectrum)
\begin{eqnarray}\label{gap}
\Delta_{fc}\sim \frac{p^2_f}{m_{fc}}\exp\left\{-\frac{2}{\epsilon}\frac{m_b}{m_{fc}}\right\}.
\end{eqnarray} 
For the precise numerical prefactor, one needs to go beyond the leading-order $\epsilon$-expansion.

\section{Vacuum trimer}\label{}
\begin{figure}[h!]
	\centerline{\includegraphics
		[width=0.475
		\textwidth,clip,angle=-0]{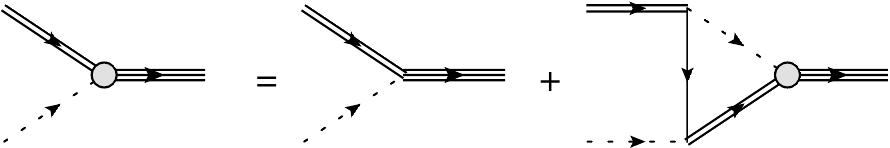}}
	\caption{The renormalized vacuum vertex function determining $b+c\to t$ process (triple line refers to trimer).}\label{bc_t_vertex_fig}
\end{figure}

\begin{figure}[h!]
	\centerline{\includegraphics
		[width=0.25
		\textwidth,clip,angle=-0]{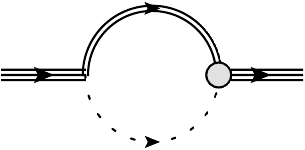}}
	\caption{The vacuum self-energy of trimer propagator.}\label{t_self_en_fig}
\end{figure}

The correct formulation of the trimer ($b-c$) problem suggests \cite{BHvK_99_1,BHvK_99_2,Hryhorchak_2023,Nakayama,Mohapatra} the modification of the original action (\ref{S}) with an extra term describing the boson-composite fermion interaction
\begin{eqnarray}\label{Delta_S}
\Delta S=-g_{cb,\Lambda}\int_xb^{\dagger}c^{\dagger}cb.
\end{eqnarray}
For later convenience let us introduce another complex Grassmann auxiliary fields that decouple composite $cb$ and $b^{\dagger}c^{\dagger}$ operators
\begin{eqnarray}\label{S_trimer}
\Delta S=g^{-1}_{t,\Lambda}\int_xt^{\dagger}t-h_{\Lambda}\int_x\left\{t^{\dagger}cb+b^{\dagger}c^{\dagger}t\right\}.
\end{eqnarray}
Clearly that $g_{cb,\Lambda}=g_{t,\Lambda}h^2_{\Lambda}$, but one cannot set $h_{\Lambda}=1$ because this vertex gets non-trivial scaling through the RG flow. A naive one-loop calculations yield
\begin{eqnarray}\label{h_Lambda}
&&\Lambda \frac{d h_{\Lambda}}{d\Lambda}=-\eta_{\textrm{RG}} h_{\Lambda}, \\ &&\eta_{\textrm{RG}}=\frac{\sin\left(\pi [d/2-1]\right)}{\pi/2}\left(\frac{m^2_c}{m_fm_t}\right)^{d/2-1},
\end{eqnarray}
with $m_t=m_c+m_b$ being mass of trimer. The solution $h_{\Lambda}=h_0\left(\frac{\Lambda}{\Lambda_0}\right)^{-\eta_{\textrm{RG}}}$ (with $\eta_{\textrm{RG}}$ always positive-definite) reflects the UV irrelevance of the appropriate coupling. Note that the above simple one-loop result is valid only in close vicinity of $d=4$, where $\eta_{\textrm{RG}}=\frac{ m^2_c}{m_fm_t}\epsilon+\dots$. The exact dependence of $h_{\Lambda}$ on UV cutoff can be extracted from the full consideration of the boson-dimer problem. Within the action, $S+\Delta S$, the possible trimer states are encoded in the vertex given by the diagrammatic equality in Fig.~\ref{bc_t_vertex_fig}. 
The appropriate integral equation for the on-shell vertex $\mathcal{T}_{bc}({\bf q})$ in the center-of-mass frame (where momenta of colliding boson and composite fermion are $\frac{m_b}{m_t}{\bf p}-{\bf q}$ and $\frac{m_c}{m_t}{\bf p}+{\bf q}$, respectively) with total momentum ${\bf p}$, reads
\begin{eqnarray}\label{Tau_bc}
&&\mathcal{T}_{bc}({\bf q})=h_{\Lambda}\nonumber\\
&&+\frac{1}{L^d}\sum_{{\bf k}}\frac{\mathcal{T}_{c}(k)\mathcal{T}_{bc}({\bf k})}{\frac{({\bf k}+{\bf q})^2}{2m_f}+\frac{k^2+q^2}{2m_b}+\frac{p^2}{2m_t}-i\nu_p},
\end{eqnarray}
with the shorthand notations for function $\mathcal{T}_{c}(k)=-\langle c^{\dagger}_Kc_K\rangle|_{i\nu_k\to i\nu_p-\frac{p^2}{2m_t}-\frac{k^2}{2m_b}}$. Note that vertex function $\mathcal{T}_{bc}({\bf q})$ also depends (although not explicitly noted) on $(d+1)$-momentum $P$, but only in combination $\frac{p^2}{2m_t}-i\nu_p$, which is consistent with the Galilean invariance of the three-body system. Before 
proceeding with the analysis of solutions to Eq.~(\ref{Tau_bc}), let us calculate the trimer propagator. It is easy to do when the on-shell vertex function $\mathcal{T}_{bc}({\bf q})$ is known. Then, the only self-energy contribution to the trimer propagator is determined by the Feynman graph shown in Fig.~\ref{t_self_en_fig}.
With the Matsubara frequency in the self-energy being explicitly integrated out, the inverse trimer propagator
\begin{eqnarray}\label{t_prop}
\langle t^{\dagger}_Pt_P\rangle^{-1}=g^{-1}_{t,\Lambda}+\frac{h_{\Lambda}}{L^d}\sum_{{\bf k}}\mathcal{T}_{c}(k)\mathcal{T}_{bc}({\bf k}),
\end{eqnarray}
is only affected by the $s$-wave channel of the boson-dimer scattering amplitude. In what follows, one needs to consider only spherically-symmetric solutions to Eq.~(\ref{Tau_bc}). In arbitrary $d$, the function $\mathcal{T}_{bc}(q)$ can be found only numerically, but the scaling limit $q\gg 1/a$ with $q\ll \Lambda$ is accessible analytically. Keeping in mind that $h_{\Lambda \to \infty}\to 0$, and that the cutoff is imposed on summation over the wave-vector in the r.h.s of (\ref{Tau_bc}), we can reduce this equation to the homogeneous one. Then, the scale invariance fixes the asymptotic form of the on-shell vertex $\mathcal{T}_{bc}(q)\propto 1/q^{\eta}$. Plugging this ansatz back into the integral equation and calculating integrals in arbitrary $d$, we arrive at the condition for allowed $\eta$s \cite{Hryhorchak_2023}
\begin{eqnarray}\label{eta}
1=-\frac{\sin(\pi d/2)}{\sin(\pi\eta/2)}
{_2}F_1\left(\frac{d-\eta}{2},\frac{d+\eta}{2}-1; \frac{d}{2};\frac{m^2_b}{m^2_c}\right).
\end{eqnarray} 
All solutions always go in pairs $\eta$ and $2-\eta$. Consequently, function $\mathcal{T}_{bc}(q)=Aq^{-\eta}+Bq^{\eta-2}$ is a linear combination of two partial solutions. Close to $d=4$ we have $\eta \approx {_2}F_1\left(2,1; 2;\frac{m^2_b}{m^2_c}\right)\epsilon =\frac{ m^2_c}{m_fm_t}\epsilon$, i.e. consistent with the above predictions $\eta_{\textrm{RG}}$ of the one-loop RG in the $\epsilon \to 0$ limit. At any particular mass ratios $m_b/m_f$, there is always range of $d\in [d_{min},d_{max}]$ (see \cite{Hryhorchak_2023}), where two mutually complex-conjugated solutions $\eta=1\pm i\eta_0$ occur. The latter is a signal for emergence of the Efimov effect in the three-body system. This is most easily seen by analyzing $\mathcal{T}_{bc}(q)\sim \frac{\sin(\eta_0\ln (q r_0))}{q}$ (here $r_0$ is arbitrary length scale) at complex $\eta$s. Being solution to homogeneous integral equation, this vertex coincides with the three-body zero-energy wave-function. The number of its nodes, therefore, determines the number of three-body bound states. Consequently, in the Efimov region we have an infinite tower of trimers with the characteristic ratio $\varepsilon_{t,n}/\varepsilon_{t,n+1}=e^{2\pi/\eta_0}$ ($n\gg 1$) between binding energies of two nearest ones. Of course, the calculations of propagators for the Efimov trimers is interesting problem, which cannot be solved to the very end in the considered model. The correct formulation requires to go beyond the zero-range boson-fermion pairwise interaction by adopting, for instance, the two-channel model describing narrow Feshbach resonance. The existence of the exact solution \cite{Gogolin_2008,Valiente_Zinner} to the three-body problem within this model in 3D inspires confidence in the calculations of propagators for the universal Efimov trimers. Apart from the interval $[d_{min},d_{max}]$, there could be at most two bound states. However, when the mass of bosonic atoms is infinite an exact solution \cite{Panochko_2021} (although only 3D case is analyzed there, the general picture is same \cite{Panochko_2022} for any $2<d<4$) of the three-body problem suggests a single bound state that satisfies bosonic statistics. Note that $m_b\gg m_f$ limit is the most favorable for the trimer emergence. Therefore, one concludes about an existence of a single trimer state with energy $\varepsilon_t\sim -\frac{1}{2m_{bc}a^2}$ outside the interval $[d_{min},d_{max}]$. With the numerically obtained vertex $\mathcal{T}_{bc}(q)$ we can calculate the trimer propagator. It is readily seen, taking into account the UV dependence of both functions $\mathcal{T}_{c}(k)\mathcal{T}_{bc}({\bf k})$, that the sum in (\ref{t_prop}) is divergent and should be renormalized. The latter is realized by absorbing the leading-order large-$\Lambda$ term in the definition of the bare coupling $g_{t,\Lambda}$. Recalling that the trimer propagator should have a simple pole in the three-body bound states, one obtains
\begin{eqnarray}\label{}
\langle t^{\dagger}_Pt_P\rangle^{-1}=\frac{h_{\Lambda}}{L^d}\sum_{{\bf k}}\left\{\mathcal{T}_{c}(k)\mathcal{T}_{bc}({\bf k})\right.\nonumber\\
\left.-\left[\mathcal{T}_{c}(k)\mathcal{T}_{bc}({\bf k})\right]_{\frac{p^2}{2m_t}-i\nu_p\to|\varepsilon_{t}|}\right\}.
\end{eqnarray}
The leading contribution to the above integral can be calculated for small $\epsilon$ utilizing an asymptotic power-law solution for the vertex $\mathcal{T}_{bc}(k)$
\begin{eqnarray}\label{}
\langle t^{\dagger}_Pt_P\rangle^{-1}\propto |\varepsilon_{t}|^{1-\eta/2}-\left[\frac{p^2}{2m_t}-i\nu_p
\right]^{1-\eta/2},
\end{eqnarray}
where the prefactor contains $h_{\Lambda}$ times constant of order unity (note that here small $\epsilon$ coming from $\langle c^{\dagger}_Pc_P\rangle$ propagator is compensated by the UV divergence of the above integral). Far below $d=4$, one needs a full solution to Eq.~(\ref{Tau_bc}) to obtain the trimer propagator. Again, exactly in $d=4$ the above expression for $\langle t^{\dagger}_Pt_P\rangle$ reproduces the propagator of a free particle with the binding energy $\varepsilon_{t}$ and mass $m_t$ of the trimer.

\section{Trimer-dimer-fermion mixture}\label{}
\begin{figure}[h!]
	\centerline{\includegraphics
		[width=0.45
		\textwidth,clip,angle=-0]{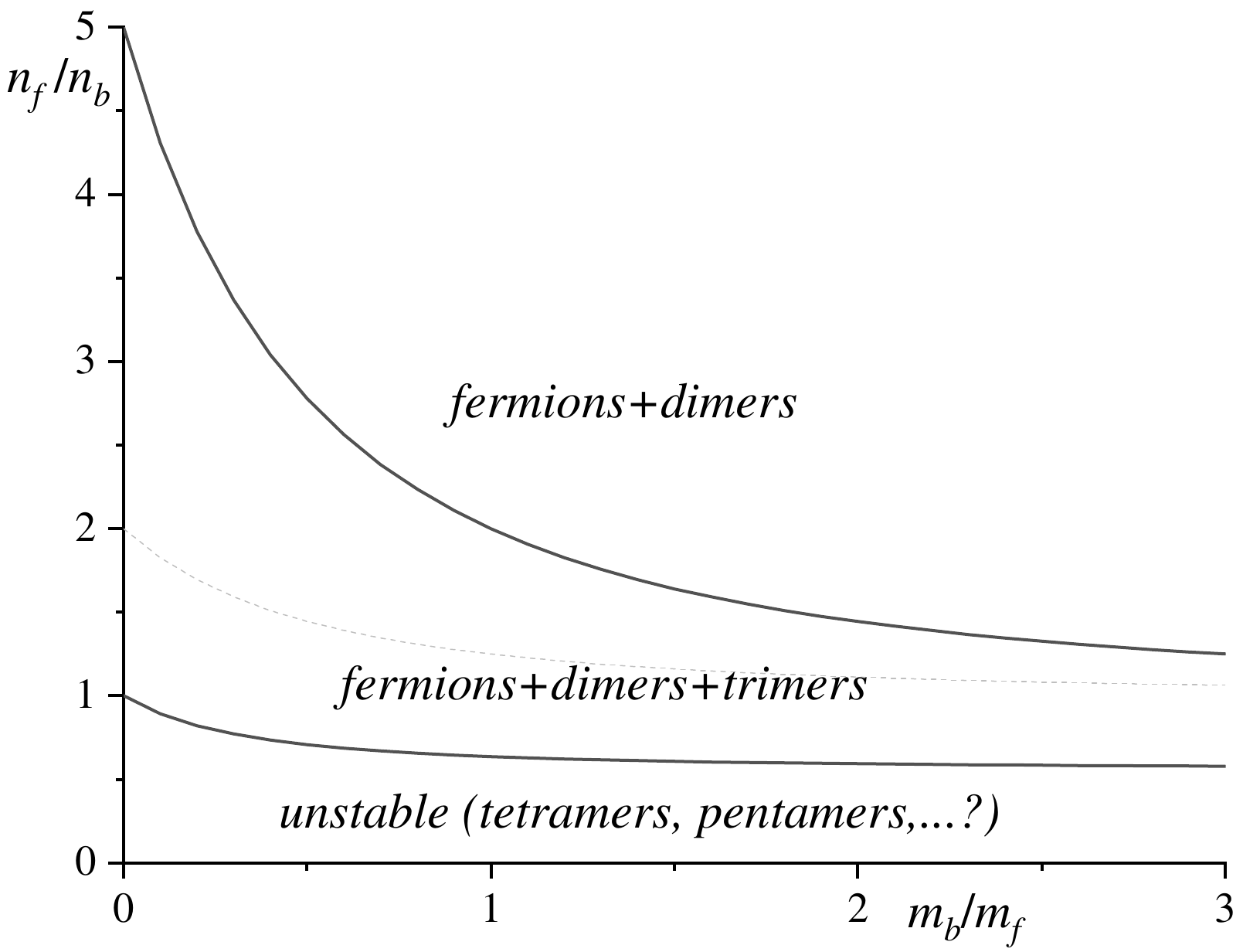}}
	\caption{The phase diagram of the four-dimensional Bose-Fermi mixture with resonant two-body interaction. Upper and lower solid lines correspond to Eq.~(\ref{dimer_trimer_trans}) and Eq.~(\ref{trimerBEC_cond}), respectively. The dashed line indicates the stability condition (\ref{dimerBEC_cond}) without the possibility for trimers to emerge.}\label{phase_diagram_fig}
\end{figure}

The existence of shallow trimers changes the above-described phase diagram of the Bose-Fermi mixture. In particular, one should expect the emergence of the thermodynamically stable phase at densities below (\ref{dimerBEC_cond}). The coexistence of trimers, dimers, and unbound fermionic atoms characterizes this phase. A simple analysis in dimensions close to $d=4$ relying on the combinations of bosonic and fermionic chemical potentials leads to the conclusion that there can be no trimers without dimers at least in the leading order of $\epsilon$ expansion. The ground state energy of this three-component Fermi mixture of almost non-interacting species can be written (up to an unimportant overall dimensionless coefficient) as follows
\begin{eqnarray}\label{}
\frac{E}{L^d}\propto \frac{(n_f-n_c-n_t)^{2/d+1}}{2m_f}+\frac{n_c^{2/d+1}}{2m_c}+\frac{n_t^{2/d+1}}{2m_t},
\end{eqnarray}
with $n_c$ and $n_t$ being densities of dimers and trimers, respectively. Their equilibrium values are subject to minimization of $E$ with the natural constraint $n_b=n_c+2n_t$ (all bosons gone for the formation of either dimers or trimers). Thus, only one density (say $n_t$) is unknown in $E$, and solving equation $\partial E/\partial n_t=0$ we find both $n_t$ and $n_c$. Then, it is easy to demonstrate that any of the solutions realize a minimum of $E$ (because $\partial^2 E/\partial n_t^2>0$). Our calculations revealed that the trimers arise below (i.e. for smaller densities of fermions $n_f$)
\begin{eqnarray}\label{dimer_trimer_trans}
\frac{n_f}{n_b}= 1+\left(\frac{2m_f}{m_c}\right)^{d/2},
\end{eqnarray}
and their number monotonically increases up to the BEC transition point, which in this case realizes at
\begin{eqnarray}\label{trimerBEC_cond}
\frac{n_f}{n_b}= \frac{m_f^{d/2}+m_c^{d/2}+m_t^{d/2}}{m_c^{d/2}+2m_t^{d/2}},
\end{eqnarray}
and for lower fermionic densities, the system becomes unstable towards collapse. The latter expression is found by writing down equations similar to (\ref{dimer_Eqs}) but with the inclusion of trimers. A maximal $n_t$ reached at (\ref{trimerBEC_cond}), reads
\begin{eqnarray}\label{}
\frac{n_t}{n_b}= \frac{m_t^{d/2}}{m_c^{d/2}+2m_t^{d/2}}.
\end{eqnarray}
It is worth stressing that the above analysis is valid only close to four dimensions. The phase diagram of the mixture exactly at $d=4$ is presented in Fig.~\ref{phase_diagram_fig}.
To reveal an order of the quantum phase transition between trimer and dimer phases (upper solid line), it is necessary to calculate the dependence of order parameter $n_t$ on the deviation of the density of fermions from critical magnitude $n^{\textrm{cr}}_f$ given by Eq.~(\ref{dimer_trimer_trans}). Fortunately, for the considered system in the $d\to 4$ limit, it is easy to do $n_t\sim (n^{\textrm{cr}}_f-n_f)^{d/2}$ (as $n_f\to n^{\textrm{cr}}_f$). Exactly at four dimensions, therefore, the system experiences a true third-order quantum phase transition (in the sense that $\partial^3 E/\partial n_f^3$ is discontinuous although finite from both sides of the critical point). In lower dimensions, the third derivative of the thermodynamic potential with respect to a control parameter is power-law divergent with a small exponent $-\epsilon/2$. 

\begin{figure}[h!]
	\centerline{\includegraphics
		[width=0.475
		\textwidth,clip,angle=-0]{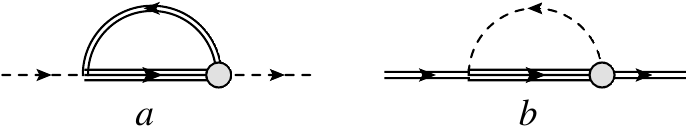}}
	\caption{Bosonic (a) and dimer (b) self-energy contributions of order $\epsilon$ due to trimers.}\label{b_self_en_2_fig}
\end{figure}

Slightly below four dimensions our system in the normal (non-superfluid) phase is a three-component weakly interacting Fermi gas. At order $\epsilon$ only bosonic self-energy (see Fig.~\ref{b_self_en_2_fig}, a)
gets corrections due to interactions with trimers. Formally, there is also a similar bubble (Fig.~\ref{b_self_en_2_fig}, b) of order $\epsilon^0$ renormalizing $\langle c^{\dagger}_Pc_P\rangle$ propagator, but effectively it contributes only to order $\epsilon^2$ to the observables. It is understood, that additional contribution to $\Sigma_c(P)$ modifies, up to order $\epsilon$, the ratio between a number of dimers and trimers in the mixture, while $\Sigma_b(K)$ shifts the BEC transition (coinciding with the system's collapse) line. The latter effect can be incorporated by using very similar formulas as in the case of dimers but with the replacement $m_f\to m_c$, $m_c\to m_t$ and $\mu_f\to \mu_c$, $\mu_c\to \mu_t$. More importantly to consider the momentum distribution $N_b(k)$ of bosons, in particular, its large-$k$ tail. The leading term (of order $1/k^4$) is a famous Tan's contribution \cite{Tan}, while the trimers provide next to the leading-order term
\begin{eqnarray}\label{N_b}
N_b(k)\sim \frac{\epsilon}{4}\left(\frac{p_c}{k}\right)^4+\frac{\eta}{4}\left(\frac{m_bm_c}{m^2_t}\right)^{\eta/2}\left(\frac{p_t}{k}\right)^{4-\epsilon+\eta}\nonumber\\
\to\frac{\epsilon}{4}\left(\frac{p_c}{k}\right)^4+\frac{\epsilon}{4}\frac{m_c^2}{m_fm_t}\left(\frac{p_t}{k}\right)^{4+\epsilon\left(\frac{m_c^2}{m_fm_t}-1\right)},
\end{eqnarray}
where $p_c$ and $p_t$ are the Fermi momenta of dimers and trimers, respectively. Note that $\frac{m_c^2}{m_fm_t}-1>0$, meaning the trimers impact is always subleading in the $k\to \infty$ limit and is different \cite{Braaten_2011} from the three-dimensional case with the intrinsic Efimov physics. Equation (\ref{N_b}) identifies a linear in $\epsilon$ terms of the two- and three-body contact parameters. Finally, moving from high temperatures (of course, $T\ll \mu_{f,c,t}$, but $T>0$) we should expect, based on previous analysis, the formation of the $p$-wave dimer-trimer $(c-t)$ Cooper pairs (in addition to the above-discussed fermion-dimer $f-c$ ones). Such a mixture then is the two-component fermionic superfluid. To reveal the variety of ground states of the system, however, one is required to perform a more detailed investigation.

\section{Summary and open questions}
In conclusion, we have studied the equilibrium properties of the strongly interacting Bose-Fermi mixture at low temperatures close to four dimensions. Assuming no interaction between bosonic atoms, we have argued that there are several thermodynamically stable states of the system without Bose-Einstein condensate. All bosons in these states are coupled to fermions, and depending on the density ratios of immersed fermions and bosons, form a macroscopic number of either dimers or dimers and trimers at once (see Fig.~\ref{phase_diagram_fig}). The latter two phases are shown to be separated by the third-order quantum phase transition. At lower densities of fermionic atoms, the BEC transition occurs signaling the instability of the fermion-dimer-trimer mixture. Once one can prove the existence of the four-body (tetramers), five-body (pentamers), and higher-order many-body bound states with the binding energies $\sim \frac{1}{m a^2}$, the system remains thermodynamic at arbitrary compositions. Although it is impossible to test these predictions experimentally, the Monte Carlo simulations can resolve many of the problems set.

Slightly below $d=4$, the thermodynamics of a system can be calculated by means of the $\epsilon=4-d$ expansion. In particular, properties of the normal phase were calculated (up to order $\epsilon$) by utilizing the Fermi liquid theory. At lower temperatures, the first superfluid phase transition should be associated with the formation of the $p$-wave Cooper pairs of fermions and dimers. In the same temperature scales the fermion-trimer pairs are also predicted constituting the considered system as a two-component superfluid. The peculiarities of the zero-temperature phase diagram, even at small $\epsilon$s, are still unknown and require further study. 

Finally, we expect that much of the presented results concerning the behavior of this unrealistic near-four-dimensional unitary Bose-Fermi mixtures should be retained in two dimensions \cite{Milczewski_2022}, especially at high densities.

\begin{center}
	{\bf Acknowledgements}
\end{center}
This work was partly supported by Project No.~0122U001514 from the Ministry of Education and Science of Ukraine.


\begin{thebibliography}{99}

\bibitem{Anderson_1995} M. H. Anderson, J. R. Ensher, M. R. Matthews, C. E. Wieman and
 E. A. Cornell, \href{https://doi.org/10.1126/science.269.5221.198}{Science {\bf 269}, 198 (1995).}
 
\bibitem{Davis_1995}  K. B. Davis, M. O. Mewes, M. R. Andrews, N. J. van Druten, D. S.
 Durfee, D. M. Kurn and W. Ketterle, \href{https://doi.org/10.1103/PhysRevLett.75.3969}{Phys. Rev. Lett. {\bf 75}, 3969 (1995).}

\bibitem{DeMarco_1999} B. DeMarco and D. S. Jin, \href{https://doi.org/10.1126/science.285.5434.1703}{Science {\bf 285}, 1703 (1999).}

\bibitem{OHara_2002} K. M. O'Hara, S. L. Hemmer, M. E. Gehm, S. R. Granade and J. E. Thomas, \href{https://doi.org/10.1126/science.1079107}{Science {\bf 298}, 2179 (2002).}

\bibitem{Navon_2011} N. Navon, S. Piatecki, K. Gunter, B. Rem, T. C. Nguyen, F. Chevy, W. Krauth, and C. Salomon, \href{https://doi.org/10.1103/PhysRevLett.107.135301}{Phys. Rev. Lett. {\bf 107}, 135301 (2011).}

\bibitem{Bedaque_2002} P. F. Bedaque U. van Kolck, \href{https://doi.org/10.1146/annurev.nucl.52.050102.090637}{Annu. Rev. Nucl. Part. Sci. {\bf 52}, 339 (2002).}

\bibitem{Braaten_2006} E. Braaten, H. W. Hammer, \href{https://doi.org/10.1016/j.physrep.2006.03.001}{Phys. Rept. {\bf 428}, 259 (2006).}

\bibitem{Wilson_1971} K. G. Wilson, \href{https://doi.org/10.1103/PhysRevB.4.3174}{Phys. Rev. B {\bf 4}, 3174 (1971);}
\href{https://doi.org/10.1103/PhysRevB.4.3184}{Phys. Rev. B {\bf 4}, 3184 (1971).}

\bibitem{Polchinski_1984} J. Polchinski, \href{https://doi.org/10.1016/0550-3213(84)90287-6}{Nucl. Phys. B {\bf 231}, 269 (1984).}



\bibitem{Stan_2004} C. A. Stan, M. W. Zwierlein, C. H. Schunck, S. M. F. Raupach, and W. Ketterle, \href{https://doi.org/10.1103/PhysRevLett.93.143001}{Phys. Rev. Lett. {\bf 93}, 143001 (2004).}

\bibitem{Inouye_2004} S. Inouye, J. Goldwin, M. L. Olsen, C. Ticknor, J. L. Bohn, and D. S. Jin, \href{https://doi.org/10.1103/PhysRevLett.93.183201}{Phys. Rev. Lett. {\bf 93}, 183201 (2004).}

\bibitem{Saam_1969} W. F. Saam, \href{https://doi.org/10.1016/0003-4916(69)90250-4}{Ann. Phys. {53}, 239 (1969).}

\bibitem{Viverit_2000} L. Viverit, C. J. Pethick, and H. Smith,, \href{https://doi.org/10.1103/PhysRevA.61.053605}{ Phys. Rev. A {\bf 61}, 053605 (2000).}

\bibitem{Powell_2005} S. Powell, S. Sachdev, and H. P. B\"uchler,
\href{https://doi.org/10.1103/PhysRevB.72.024534}{Phys. Rev. B {\bf 72}, 024534 (2005).}

\bibitem{Watanabe_2008} T. Watanabe, T. Suzuki, and P. Schuck, \href{https://doi.org/10.1103/PhysRevA.78.033601}{Phys. Rev. A
	{78}, 033601 (2008).}

\bibitem{Marchetti_2008} F. M. Marchetti, C. J. M. Mathy, David A. Huse, and M. M. Parish, \href{https://doi.org/10.1103/PhysRevB.78.134517}{Phys. Rev. B {\bf 78}, 134517 (2008).}

\bibitem{Fratini_2012} E. Fratini and P. Pieri, \href{https://doi.org/10.1103/PhysRevA.85.063618}{Phys. Rev. A {\bf 85}, 063618 (2012).}

\bibitem{Hryhorchak_2023} O. Hryhorchak, V. Pastukhov, \href{https://doi.org/10.1088/1751-8121/accda4}{J. Phys. A: Math. Theor. {\bf 56}, 205003 (2023).}

\bibitem{Fratini_2010} E. Fratini and P. Pieri, \href{https://doi.org/10.1103/PhysRevA.81.051605}{Phys. Rev. A {\bf 81}, 051605(R) (2010).}

\bibitem{Manabe_2019} K. Manabe, D. Inotani, and Y. Ohashi, \href{https://doi.org/10.1103/PhysRevA.100.063609}{Phys. Rev. A {\bf 100}, 063609 (2019).}

\bibitem{Guidini_2015} A. Guidini, G. Bertaina, D. E. Galli, and P. Pieri, \href{https://doi.org/10.1103/PhysRevA.91.023603}{Phys. Rev. A {\bf 91}, 023603 (2015).}

\bibitem{Ludwig_2011} D. Ludwig, S. Floerchinger, S. Moroz, and C. Wetterich, \href{https://doi.org/10.1103/PhysRevA.84.033629}{Phys. Rev. A {\bf 84}, 033629 (2011).}

\bibitem{Bertaina_2013} G. Bertaina, E. Fratini, S. Giorgini, and P. Pieri, \href{https://doi.org/10.1103/PhysRevLett.110.115303}{Phys. Rev. Lett. {\bf 110}, 115303 (2013).}

\bibitem{Duda_2023} M. Duda, X.-Y. Chen, A. Schindewolf, R. Bause, J. von
Milczewski, R. Schmidt, I. Bloch, and X.-Y. Luo, \href{https://doi.org/10.1038/s41567-023-01948-1}{Nat. Phys. {\bf 19}, 720 (2023).}

\bibitem{Guo_2023} Y. Guo, H. Tajima, T. Hatsuda, and H. Liang, \href{https://doi.org/10.1103/PhysRevA.108.023304}{Phys. Rev. A {\bf 108}, 023304 (2023).}


\bibitem{Park_2012} J. W. Park, C.-H. Wu, I. Santiago, T. G. Tiecke, S. Will, P. Ahmadi, and M. W. Zwierlein, \href{https://doi.org/10.1103/PhysRevA.85.051602}{Phys. Rev. A {\bf 85}, 051602(R) (2012).}

\bibitem{Heo_2012} M.-S. Heo, T. T. Wang, C. A. Christensen, T. M. Rvachov, D. A. Cotta, J.-H. Choi, Y.-R. Lee, and W. Ketterle, \href{https://doi.org/10.1103/PhysRevA.86.021602}{Phys. Rev. A {\bf 86}, 021602(R) (2012).}

\bibitem{Cumby_2013} T. D. Cumby, R. A. Shewmon, M.-G. Hu, J. D. Perreault, and D. S. Jin, \href{https://doi.org/10.1103/PhysRevA.87.012703}{Phys. Rev. A {\bf 87}, 012703 (2013).}

\bibitem{Cao_2023} J. Cao, H. Yang, Z. Su, X.-Y. Wang, J. Rui, B. Zhao, and J.-W. Pan, \href{https://doi.org/10.1103/PhysRevA.85.051602}{Phys. Rev. A {\bf 107}, 013307 (2023).}

\bibitem{Yang_2022} H. Yang, J. Cao, Z. Su, J. Rui, B. Zhao, J.-W. Pan, \href{https://doi.org/10.1126/science.ade6307}{Science {\bf 378}, 1009 (2022).}

\bibitem{Efimov} V. Efimov, \href{https://doi.org/10.1016/0370-2693(70)90349-7}{Phys. Lett. B {\bf 33}, 563 (1970).}

\bibitem{Naidon} P. Naidon and S. Endo, \href{https://doi.org/10.1088/1361-6633/aa50e8}{Rep. Prog. Phys. {\bf 80}, 056001 (2017).} 

\bibitem{Rosa_2018} D. S. Rosa, T. Frederico, G. Krein, and M. T. Yamashita, \href{https://doi.org/10.1103/PhysRevA.97.050701}{Phys. Rev. A {\bf 97}, 050701(R) (2018).}


\bibitem{Nussinov_2006} Z. Nussinov and S. Nussinov, \href{https://doi.org/10.1103/PhysRevA.74.053622}{Phys. Rev. A {\bf 74}, 053622 (2006).}

\bibitem{Nishida_2012} Y. Nishida, D.T. Son, \href{https://doi.org/10.1007/978-3-642-21978-8_7}{Unitary Fermi Gas, 
	Expansion, and Nonrelativistic Conformal Field Theories. In:  W. Zwerger,(eds) The BCS-BEC Crossover and the Unitary Fermi Gas. Lecture Notes in Physics, vol 836. Springer, Berlin, Heidelberg (2012).}

\bibitem{Nishida_2006} Y. Nishida and D. T. Son, \href{https://doi.org/10.1103/PhysRevLett.97.050403}{Phys. Rev. Lett. {\bf 97}, 050403 (2006).}



\bibitem{Brodsky_2006} I. V. Brodsky, M. Yu. Kagan, A. V. Klaptsov, R. Combescot, and X. Leyronas, \href{https://doi.org/10.1103/PhysRevA.73.032724}{Phys. Rev. A {\bf 73}, 032724 (2006).}

\bibitem{Levinsen_2011} J. Levinsen,  D. S. Petrov, \href{https://doi.org/10.1140/epjd/e2011-20071-x}{Eur. Phys. J. D {\bf 65}, 67 (2011).}

\bibitem{Abramowitz} M. Abramowitz and I. Stegun, {\it Handbook of Mathematical
	Functions with Formulas, Graphs, and Mathematical Tables} (United States Department of Commerce, National Bureau of Standards 1964).

\bibitem{BHvK_99_1}  P. F. Bedaque, H.-W. Hammer, and U. van Kolck, \href{https://doi.org/10.1103/PhysRevLett.82.463}{Phys. Rev. Lett. {\bf 82}, 463 (1999).}

\bibitem{BHvK_99_2} P. F. Bedaque, H.-W. Hammer, and U. van Kolck, \href{https://doi.org/10.1016/S0375-9474(98)00650-2}{Nucl. Phys. A {\bf 646}, 444 (1999).}

\bibitem{Mohapatra} A. Mohapatra and E. Braaten \href{https://doi.org/10.1103/PhysRevA.98.013633}{Phys. Rev. A {\bf 98}, 013633 (2018).}

\bibitem{Nakayama} Y. Nakayama and Y. Nishida, \href{https://doi.org/10.1103/PhysRevE.103.012117}{Phys.~Rev.~E {\bf 103}, 012117 (2021).}

\bibitem{Gogolin_2008} A. O. Gogolin, C. Mora, and R. Egger, \href{https://doi.org/10.1103/PhysRevLett.100.140404}{
Phys. Rev. Lett. {\bf 100}, 140404 (2008).}

\bibitem{Valiente_Zinner} M. Valiente, and N. T. Zinner, \href{https://dx.doi.org/10.1088/978-0-7503-3087-9}{{\it Strongly Interacting Quantum Systems}, Volume 1 (IOP Publishing 2023).} 

\bibitem{Panochko_2021} G. Panochko and V. Pastukhov, \href{https://doi.org/10.1088/1751-8121/abdbc5}{J. Phys. A: Math. Theor. {\bf 54}, 085001 (2021).}

\bibitem{Panochko_2022} G. Panochko and V. Pastukhov, \href{https://doi.org/10.3390/atoms10010019}{Atoms {\bf 10}, 19 (2022).}

\bibitem{Tan} S. Tan, \href{https://doi.org/10.1016/j.aop.2008.03.005}{Ann.~Phys. {\bf 323}, 2971 (2008).}

\bibitem{Braaten_2011} E. Braaten, D. Kang, and L. Platter, \href{https://doi.org/10.1103/PhysRevLett.106.153005}{Phys. Rev. Lett. {\bf 106}, 153005 (2011).}


\bibitem{Milczewski_2022} J. von Milczewski, F. Rose, R. Schmidt, \href{https://doi.org/10.1103/PhysRevA.105.013317}{Phys. Rev. A {\bf 105}, 013317 (2022).}


\end{thebibliography}
\end{document}